# Smart Societies: From Citizens as Sensors to Collective Action

*Andrés Monroy-Hernández, Shelly Farnham, Emre Kıcıman, Scott Counts, Munmun De Choudhury*

As the scale of social media progresses to global ubiquity, an increasing number of fields of study have explored the intersection of *computing and society*. We see this in areas as disparate as political science, health, finance, sociology, and urban planning. Social computing touches nearly every aspect of the human experience. For example, social media are so pervasive that the societal implications of sociotechnical systems have become part of everyday conversations, from news coverage of "Twitter revolutions" to recent debates around the potentials and tribulations of Big Data, in the research community, mainstream media, and general public alike.

In the past decade, one of the most striking transformations in people's everyday experiences with technology is that computing has extended its reach *beyond the home and the enterprise* and *into the town square*. Interactive systems are no longer limited to the realm of productivity and entertainment, they are becoming an integral part of *public discourse* and *collective action*.

Early on in the history of computing, productivity software became prevalent at home and in enterprise focusing on "getting things done," often in the context of private, top-down organizations. A few years later, creative and playful software reached the mainstream. Today, social computing has emerged as the new paradigm (See Figure 1).

Social computing has augmented existing connections among friends and family, but perhaps more importantly, it has facilitated interactions among people with weak links, or no previous links at all, leading to this decade's paradigm of societal level computing.

Societal computing embodies a vision of computing as an instrument for gaining awareness of *societal patterns*, and for supporting large-scale *collective action* where people collaborate to achieve common social goods. We see this often in the context of public, bottom-up, agile, ad-hoc organizations at the level of neighborhoods, cities, countries, and even global communities.

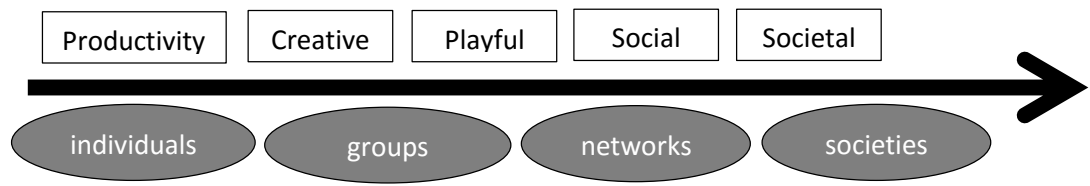

**Figure 1: Computing paradigms**

In this article, we focus on two areas of societal computing that we think should drive its research and development: societal pattern discovery, and collective action. We close with reflections on the key challenges we and others have identified such as issues around privacy, biases, and collectivism.

**Discovering Latent Societal Patterns**

The ubiquitous computing devices embedded in buildings and appliances play the role of sensors collecting physical data. Similarly, the social computing systems that mediate our everyday interactions with other people act as *social sensors* that collect large amounts of social data. These data, although biased and not all-inclusive, help us probe our collective attitudes and behaviors.

For example, recent studies have shown that it is possible to model and predict health patterns using existing social sensors. More specifically, using data from Twitter, researchers have modeled the epidemics of colds in New York City [9] and post-partum mood changes among new mothers [3]. Similarly, these systems have been used to model daily lifestyles [5], urban dynamics [4], elections [14], and even disasters [2].

Social media analytics is a fairly nascent field, with newly emerging tools and best practices for processing big data, information visualization, sentiment analysis, and so forth. Research and applications in this space require multidisciplinary skills ranging from machine learning, social network analysis, log analysis, content analysis, distributed data systems, visualization design, as well as the domain-specific knowledge in fields such as sociology, public health, or urban planning. Merging computer and social science can be difficult, but this is the combination that will move societal computing forward in meaningful ways.

Examining large-scale social data helps us understand and reflect on *historical and real-time aggregate behavior and opinions* by uncovering trends that recency bias would otherwise obscure, or that are hard to parse as events unfold. This information analysis is useful not only to governments and large organizations but also to average citizens who can gain a deeper understanding of their society. These technologies can enable smarter, healthier societies, including meta-reflection on the role of the technologies themselves, and eventually lead to collective and individual action.

**Supporting Collective Action**

Beyond helping people understand societal patterns, sociotechnical systems can encourage, or help support, collective action and social change.

Among those with access and skills, the participatory culture of social media affords people with new opportunities to connect with civic life, by enabling even disenfranchised individuals to express their voices and connect with similar others.

The ubiquitous integration of communication and social media technologies in people's everyday lives has made it possible for people's individual voices to coalesce and interact in the networked public sphere. This change has transformed how social mobilizations occur, for example, during the 2010 U.S. congressional election a 61-million-person experiment found that Facebook messages "directly influenced political self-expression, information seeking and real-world voting behavior of millions of people" [1]. Furthermore, researchers studying the 2011 Egyptian revolution found that Facebook and Twitter increased people's likelihood to attend the biggest demonstration in Tahir Square [13].

These new civic media tools, platforms, channels, and practices facilitate civic engagement --those civic and political activities motivated by a desire for social change-- ranging from community awareness to voting, from government accountability to public dissent. Civic media is motivated by a desire to foster social change in both local and global communities. Beyond politics, people have used these civic media technologies for collective action in the wake of acute events such as floods [11], earthquakes [12], and even wars [7].

The low effort, participatory, networked communication of social media now enables a variety of organizational structures (informal, networked, and individual) to achieve the key tasks of collective action heretofore largely the domain of formal organizations: *identifying people* with common interests, *communicating* with them as a group, and *coordinating* their efforts.

The diversity of civic activities supported by social media suggests a need for a cohesive framework for the design of civic media systems. While the processes of collective action and civic engagement have been well-studied within fields such as political science and sociology, there is much work to be done in our fields of HCI, CSCW, and Social Computing, to develop a theoretical framework and a set of system design patterns for *computer-supported collective action*.

## Challenges: Privacy, Biases, and Collectivism

Several people have problematized and voiced skepticism about the utopianism that often surrounds technology's role in society [6, 8]. The issues often stem from society's power inequity, from the power that comes from information when it comes to privacy, to the power people with access and skills to socio-technologies have over those who do not, to the power that crowds can misuse.

Among the issues raised with regards to big data, the topic of *privacy* comes up frequently. For example, when using social media to investigate public health patterns, questions about the nature of public and personal health data raise questions as different sources of data carry different assumptions of privacy and the possibility of disclosing sensitive information inadvertently. Similarly, when it comes to civic media, the loss of privacy at the hand of oppressive regimes or criminal organizations can put people's lives in danger, while in the context of rich societies it can help give more power to existing corporate interests. Therefore, an important issue in the design of these systems is how to enable people to negotiate their public identity in a way that also meets their privacy needs.

In addition, one of the issues with many technological-based approaches to addressing societal problems is that of *bias* in usage, in terms of demographics, topics. For example, several studies have shown gaps in technological access and skills based on race and income, as well as the wide variety of biases the data might have. One instance of this failure to represent the real world was the inaccuracy of Google Flu predictions in 2012, despite apparent success in previous years. An important area for research is understanding the nature of these biases, and to develop best practices in controlling for them in our research methods and applications.

Lastly, one of the issues with empowerment is that it can lead to collectivism: as responsibility is diffused among the crowd, witch-hunts, false rumors, and other unsavory outcomes can occur. For example, following the recent Boston Marathon bombing, a group of eager amateur sleuths, carefully picking through the images and videos made available through social media, had come to the wrong conclusion about one of the perpetrators. How then, can we foster a culture of accurate citizen journalism, develop tools for identifying false or unreliable information, and discourage misguided mob behaviors?

These are only a few of the challenges that must be addressed when designing for effective civic media, yet we believe they are a necessary part of our ongoing conversation on the development of new societal computing technologies. As a community, it is our task to explore how to best harness the powerful advantages of this new age of big social data and participatory social media toward a smarter, more engaged civic society, while also mitigating potential negative outcomes. To date, we have collected immense amounts of social data, and observed striking changes in the evolution of social movements. However, we are only beginning to scratch the surface in converging on best practices for large scale social media analytics, and in developing the tools for more deliberative, effective collective action.

**About the Authors**

Andrés Monroy-Hernández, Shelly Farnham, Emre Kiciman, Scott Counts, and Munmun De Choudhury are researchers at Microsoft Research working on social media analytics and civic media.